# Positioning of self-assembled Ge islands on stripe-patterned Si (001) substrates

Zhenyang Zhong, A.Halilovic, M. Mühlberger, F. Schäffler, G. Bauer

*Institute for Semiconductor Physics, Johannes Kepler University,*
*A-4040 Linz, Austria*

**Abstract:**

Self-assembled Ge islands were grown on stripe-patterned Si (001) substrates by solid source molecular beam epitaxy. The surface morphology obtained by atomic force microscopy (AFM) and cross-sectional transmission electron microscopy images (TEM) shows that the Ge islands are preferentially grown at the sidewalls of pure Si stripes along [-110] direction at 650° C or along the trenches, whereas most of the Ge islands are formed on the top terrace when the patterned stripes are covered by a strained GeSi buffer layer. Reducing the growth temperature to 600°C results in a nucleation of Ge islands both on the top terrace *and* at the sidewall of pure Si stripes. A qualitative analysis, based on the growth kinetics, demonstrates that the step structure of the stripes, the external strain field and the local critical wetting layer thickness for the islands formation contribute to the preferential positioning of Ge islands on the stripes.

## I. INTRODUCTION

Self-assembled quantum dots (SAQD) have become an intensive research topic not only because of their promising device applications[1] but also to understand the fundamental process of strained thin film growth. A frequently employed growth mode of a strained heteroepilayer is the so-called Stranski-Krastanow(SK) mode, where after a thin wetting layer the epilayer releases the misfit strain by 3D island formation. These straightforwardly formed islands can be small and in general dislocation-free, but random in position. Although locally laterally ordered SAQD have been observed, i.e., via the growth of a multilayer of SAQD,[2,3] or the growth of SAQD above a buried strained layer with a dislocation network,[4] or growth of SAQD on a vicinal surface with step-bunching,[5,6] these kinds of short-range ordered SAQD are not adequate for most electronic or optoelectronic device applications.[1] Recently, it was shown that the combination of lithographic etching techniques and the SK growth mode provides a potential to grow long-range spatially ordered SAQD.[7-13] Some interesting phenomena have been observed during the growth of SAQD on patterned substrates. For instance InAs SAQD can be preferentially grown either on the top terraces or at the sidewalls or at the trenches of stripe-patterned GaAs substrates.[7] The number of InAs SAQD can be adjusted by the depth of patterned holes.[8] Ge SAQD are preferentially grown at the edges of selectively grown Si mesas in etched $SiO_2$ windows,[9,10] while they preferentially grow at the sidewalls of pure Si stripes.[11] These phenomena were discussed from the point of view of either energetics or kinetics. However, no detailed discussion about the growth of SAQD on patterned substrates has been presented so far.

It is the purpose of this paper to present a description of a growth mechanism to explain the main features of the Ge island formation on stripe-patterned Si (001) substrates. The preferential positioning of Ge SAQD grown at 650° C on patterned pure Si stripes with different periodicity and height is analyzed from AFM images and cross-sectional TEM images. The importance of the growth temperature for the preferential nucleation of Ge islands on patterned stripes is observed and discussed. In addition, we obtained the strain distribution at the surface of the stripes with a SiGe strained layer by a finite element method (FEM) simulation. The effect of this external strain field on the island formation is then discussed. Our results suggest that primarily the growth kinetics affects the preferential positioning of SAQD on the patterned substrates.

## II. EXPERIMENT

The samples were grown by solid source molecular beam epitaxy in a Riber SIVA 45.[14] During growth, the background pressure is about $10^{-10}$ Torr. The temperature was measured by a thermo couple which has been calibrated to $\pm 15^{\circ}C$.[14] The patterned stripes along <110> direction with a period of less than $1 \mu m$ were fabricated by holographic lithography and reactive ion etching (RIE) on high resistivity (>1000 $\Omega cm$) p-type Si (001) substrates. In order to study the influence of (i) the sidewall angle and the widths of the sidewalls on the nucleation of islands, a series of patterned substrates was fabricated. After etching all samples were overgrown with a Si buffer. Furthermore, prior to deposition of pure Ge in order to initiate island formation, some of the patterned samples were overgrown with GeSi to study the influence of inhomogeneous strain fields on the island nucleation. The periodicities of the stripes and the depths of the trenches after RIE of the samples are listed in Table I. The pre-patterned substrates were cleaned by a RCA process without HF dip (for samples X1-X4) or with a final HF dip (for samples X6, X7, $X_T$) before they were put into the load-lock chamber of the MBE apparatus. The oxide layer was desorbed at 900 $^{\circ}C$ for three minutes.

For samples X1 and X2 with trenches of about 1000 Å deep, a total 1330Å thick Si buffer layer was deposited, followed by 7 monolayer (ML) Ge deposition at 650 $^{\circ}C$ at the rate of 0.1Å/s with subsequent annealing at 650 $^{\circ}C$ for 35 seconds. For sample $X_T$, after the same buffer layer growth, 7 ML Ge were deposited at 600$^{\circ}$ C with subsequent annealing at 600$^{\circ}$ C for 1 minute. Similar growth processes were used to grow sample X3 and X4, except that a superlattice consisting of 5 periods of 20Å $Si_{0.5}Ge_{0.5}$ / 30Å Si grown at 550 $^{\circ}C$ was inserted into the buffer layer. Further growth details for samples X1-X4 can be found in Ref.11. For samples X6 and X7, due to the shallow depth of the patterned stripes (about 500 Å), a total of 1000 Å Si was grown at 0.5Å/s, while the substrate temperature was increased from 550 $^{\circ}C$ to 650 $^{\circ}C$. 7ML Ge was then deposited at 0.05Å/s with subsequent annealing at 650$^{\circ}$ C for 70s. After growth, the substrate temperature was decreased rapidly to room temperature. Detailed information on the layer growth is listed in Table I.

The surface morphologies of the samples were measured after growth in air by a Park Scientific atom-force microscope (AFM). The scan direction was perpendicular to the stripes. The cross-sectional TEM images were obtained in a Jeol 2011 at 200 keV after a standard dimple/Ar-ion milling preparation.

## III. RESULTS AND DISCUSSION

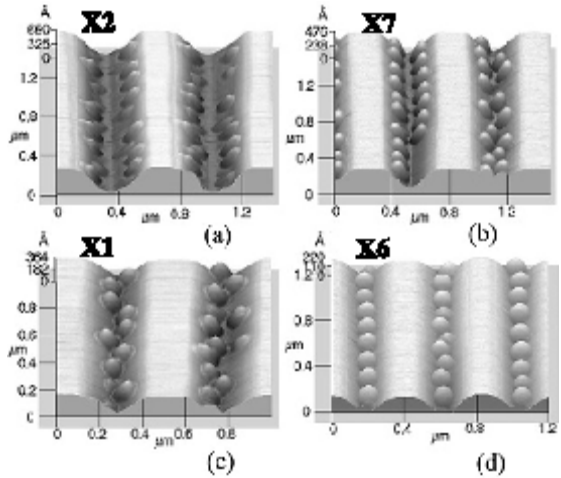

**Fig. 1** AFM micrographs of Ge islands formed on stripe-patterned Si substrates of (a) sample X2, (b) sample X7, (c) sample X1, (d) sample X6.

Figure 1 shows the AFM images of sample X2, X7, X1 and X6 where a pure Si buffer was deposited on the patterned stripes prior to Ge growth. Apparently, no Ge islands are grown on the top terraces of the stripes in these samples. In addition, most Ge islands at the sidewalls are at random positions, as shown in Fig. 1 (a)-(c). With a smaller period and a shallow depth of the trenches we achieve indeed arrays of one-dimensionally ordered islands at the bottom of the trenches, as shown in Fig. 1(d). The root-mean-square (rms) roughness of the sample surface on the top terraces is about 1.3Å and 2.4Å perpendicular and along the stripes, respectively. Due to the limited resolution of AFM, the appearance of the islands and the rather steep sidewalls, the accurate rms roughness value at the sidewalls cannot be obtained. Nevertheless it should be similar to that on the top terraces.

**Table I.** The period of the stripes, the depth after etching, the buffer layer thickness, Ge growth rate and growth temperature T, the top terrace width of the stripes, the depth of the trenches after growth and the steepness of the middle part of the sidewall for all samples.

| Sample | X2 | X7 | X1 | X6 | X4 | X3 | $X_T$ |
|---|---|---|---|---|---|---|---|
| Period (Å) | 6700 | 6000 | 5000 | 4000 | 7500 | 5000 | 5000 |
| Etching depth (Å) | ~1000 | ~500 | ~1000 | ~500 | ~1000 | ~1000 | ~500 |
| buffer layer thickness (Å) | 1330 | 1000 | 1330 | 1000 | 1330[a] | 1330[a] | 1330 |
| Ge growth rate (Å/s) | 0.1 | 0.05 | 0.1 | 0.05 | 0.1 | 0.1 | 0.1 |
| Ge growth T (°C) | 650 | 650 | 650 | 650 | 650 | 650 | 600 |
| Width (Å) | 1500 | 2100 | 1750 | 1100 | 1300 | 1980 | 2220 |
| Depth (Å) | 566 | 353 | 280 | 181 | 895 | 252 | 173 |
| Steepness (°) | 21.5 | 15.3 | 11.8 | 9.3 | 26.0 | 13.2 | 10.2 |

[a] for sample X4 and X3, a five period 250 Å thick $Si_{0.5}Ge_{0.5}/Si$ (20 Å /30 Å) superlattice layer is inserted after 1000Å Si buffer layer growth, and additional 80Å Si layer is deposited before Ge deposition.

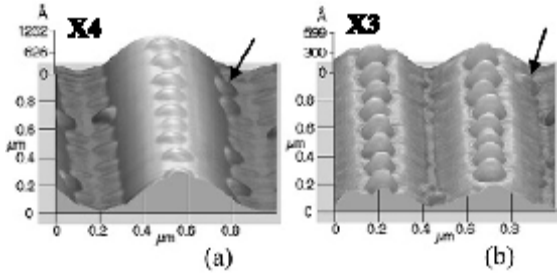

**Fig. 2** As Fig.1 but for stripe patterned Si substrates with a deposition of a Si/SiGe superlattice prior to Ge island growth: (a) sample X4, (f) sample X3. The black arrows in (a) and (b) indicate that some islands also nucleate at the bottom part of the sidewalls.

On the other hand, for samples X4 and X3 with a strained Si/SiGe superlattice buffer layer, as shown in Fig. 2, Ge islands are mainly grown on the top terrace of the stripes, forming there one-dimensionally ordered arrays. In addition some islands (denoted by a black arrow) nucleate near the bottom of the sidewalls, as shown in Figs. 2 (a) and (b). It is promising that via the patterning and growth sequences described above one-dimensionally ordered islands can be achieved on stripe-patterned substrate, either at the bottom of the trenches or on the top terrace of the stripes, as shown in Fig. 1 (d), Fig. 2(a) and (b), by changing either the shape of the trench or the strain status of the buffer layer. Details about the geometry of the stripes, as obtained from the AFM images, are listed in Table I.

Cross-sectional TEM images of sample X7, X6 and X3 are shown in Fig. 3 (a), (b) and (c), respectively. Ge islands are observed at the sidewall of the stripes or at the bottom of the trenches in sample X7 and X6, as denoted by small black arrows in Figs. 3 (a) and (b). However, for sample X3 with a strained buffer layer, the Ge atoms reside on the top terrace to form islands there, whereas only few Ge islands are formed at the sidewalls, as denoted by the small and large black arrows in Fig. 3 (c). These observations agree with those from the AFM images. Another important feature of these TEM images is the absence of dislocations. In Figs. 3 (a) and (b), a black dotted line, as indicated by a white arrow, can be seen. This contrast stems from small SiC clusters induced by the additional HF dip after the RCA cleaning process for samples X6 and X7. This SiC contamination can be regarded as a good marker to indicate the position of the interface between the buffer layer and the substrate, exhibiting the cross-sectional Si profile of the stripes before the buffer layer growth. More interestingly, it reveals the actual position of the Ge islands on the surface with respect to the buried stripe pattern. The effect of these SiC clusters on the Ge islands formation can be ignored because they are small in comparison to the buffer layer thickness, dislocation-free, and coherently overgrown under the growth condition applied here. For sample X3, due to the HF-free RCA cleaning process and the subsequent complete thermal desorption of the native $SiO_2$ layer, nearly no Carbon-contamination exists. Therefore, the interface between the buffer layer and the substrate cannot be seen in Fig 3. (c). However, the strained superlattice layers can be clearly distinguished from their surroundings throughout the entire structures.

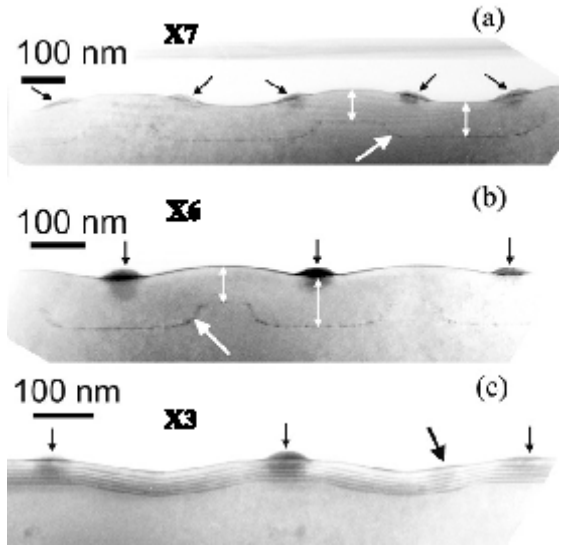

**Fig. 3** Cross-sectional TEM images of (a) sample X7, (b) sample X6, (c) sample X3. The black arrows indicate the Ge islands. In (a) and (b), the large white arrows indicate the interface between the buffer layer (marked by SiC) and the substrate, the small white two-way arrows indicate the thickness of the buffer layer.

Our observations demonstrate that Ge islands are not preferentially grown at the edges of the top terraces of the stripes. STM investigations[15,16] of a sub-monolayer Ge deposition on Si (001) substrates have also demonstrated that Ge/Si intermixing occurs randomly on the terraces and not preferentially at steps or points defects. This means that the growth kinetics, i.e. the migration of adatoms, will significantly affect the nucleation and formation of Ge SAQDs. It has been found that the migrating units on the reconstructed surface are essentially the ad-dimers.[17,18] Therefore, in the following discussion, we will focus on the migration of ad-dimers on the surface rather than monomers. During Ge deposition, the ad-dimers are either Ge ad-dimers or Ge-Si ad-dimers. The Ge-Si ad-dimers mainly appear at the beginning of Ge deposition.[17]

Before interpreting the migration of ad-dimers on the patterned substrate the type of the steps at the sidewall of the stripes should be taken into account. Considering the surface reconstruction on the terraces and the height of the steps, there are generally four types of steps on the surface, single layer steps with the edge parallel ($S_A$) or perpendicular ($S_B$) to the dimer row of the upper terrace, and double layer steps with the edge parallel ($D_A$) or perpendicular ($D_B$) to the dimer row of the upper terrace. It has been found that the $D_B$ step is energetically more favorable than the mixture of $S_A$ and $S_B$ steps, when the contact angle of the vicinal surface is larger than a critical angle.[19,20] This has been confirmed by STM observations.[16] *Ab initio* studies[21] demonstrated that the $S_B$ step grows faster than the $S_A$ step, resulting in the formation of the $D_B$ step. Oshiyama[20] has suggested that the {311} and {111} facets microscopically originate from $D_B$ step bunching. Therefore, we assume that the sidewalls of the stripes along <110> direction are mainly composed of $D_B$ steps in our samples.

Firstly, the migration of the ad-dimers on the surface of the patterned stripes without any external strain field is discussed. Provided that a vacancy in the terrace can diffuse to the step,[16] the top terrace of the stripe in our samples is regarded as vacancy-free in comparison to the plain surface. Therefore, at the beginning of each monolayer growth, the mean time for ad-dimers on the top terrace migrating to one edge, $t_d$, can be estimated by the following expressions,

$$t_d = l_1 l_2 / D, \quad D = a^2 f, \quad f = \gamma \exp(-E_b / k_B T) \quad (1)$$

where $l_1$ and $l_2$ ($l_1+l_2=w$, width of the top terrace) are the distances between the original ad-dimer position and the two edges of the top terrace of the stripes, respectively; $D$ is the diffusion constant, $a$ is the surface lattice constant (3.84Å), $f$ is the hopping rate, $\gamma$ (~$10^{13}$) is the prefactor, $k_B$ is the Boltzmann constant, $T$ denotes the growth temperature, $E_b$ is the diffusion barrier. The maximum mean migration time, $t_{dmax}$, is for ad-dimers at the center of the top terrace. The wider the top terrace, the longer $t_{dmax}$. The mean width of the top terrace in samples X1, X2, X6 and X7 is smaller than 2100Å. Therefore, at 650° C, $t_{dmax}$ for Ge-Si ad-dimers is estimated to be less than 2.5ms assuming a diffusion barrier of 1.01eV.[17] On the other hand, if we assume that a triple-atom static cluster can be formed when the migrating ad-dimers meet an adatom on their way to the edge of the top terrace, the mean nucleation time, $t_n$, of this small static cluster can be calculated by the following equation,

$$t_n = \left(\frac{a_{Ge}}{4g}\right) \bigg/ \left(\frac{3w}{a}\right) \qquad (2)$$

where $a_{Ge}$ is the bulk lattice constant of Ge, $g$ is the growth rate. The factor of 3 appears because not only the adatom deposited on the same row of an atomic site on the surface with the ad-dimer can trigger the nucleation of the small cluster, but also those on the both neighboring rows. The wider the top terrace, the shorter $t_n$. For the growth rate of 0.1 Å/s, we obtain $t_n \geq 8.6 ms$ for samples X1, X2, X6 and X7. Apparently, $t_{dmax}$ (about 2.5ms for SiGe ad-dimers and even smaller for Ge ad-dimers) is much smaller than $t_n$. This means that the Ge-Si (or Ge) ad-dimers can readily migrate to the edge of the top terrace at the beginning of each monolayer growth.

A part of these ad-dimers at the edge of the terraces can then migrate downward over the steps. Theoretical calculations[21,22] and experimental results[23] have demonstrated that the activation barrier at the steps for Si ad-dimers (or adatoms) migrating downward is smaller than upward. Due to very similar chemical and electronic properties of Si and Ge, a similar asymmetric activation barrier for Ge-Si (or Ge ) ad-dimers at the steps is assumed. This means that more ad-dimers can migrate downwards than upwards at the steps. As a result, a net flux of Ge-Si (or Ge) ad-dimers exists, migrating downwards from the top terrace to the bottom of the sidewall. Moreover, the amount of ad-dimers of this net downward flux can be quite large at the growth temperature of 650° C. The activation barrier for Si ad-dimers migrating downward at $D_B$ step, $E_{bs}$, is calculated to be 1.55 eV.[22] Considering that the activation barrier for Ge-Si[17] and Ge[24] ad-dimers on the dimer rows is smaller by about 0.1 eV and 0.3 eV than that for Si ad-dimers[18], a simple estimate of the activation barrier at $D_B$ for Ge-Si and Ge ad-dimers yields 1.45 and 1.25 eV, respectively. The hopping-down rates of Ge-Si and Ge ad-dimers at $D_B$ steps are then estimated to be $1.2 \times 10^5$ and $1.5 \times 10^6$/s by replacing $E_b$ in Eq. (1) by the $E_{bs}$, respectively. Recent experimental results[23] indicate that the activation barriers at steps may be even smaller than the theoretical values[22]. So the hopping rates of ad-dimers at the steps should then be even larger. The other factor related to the downward flux of ad-dimers is the magnitude of the asymmetric activation barriers at steps. This activation barrier difference at $D_B$ is about 0.3 eV [22] for Si ad-dimers. Considering the similarities between Ge and Si and the weaker bond strength of Ge-Ge and Ge-Si with respect to that of Si-Si, this barrier difference for a Ge (or Ge-Si) ad-dimer is of the same order but smaller. Assuming a difference of 0.1eV, the hopping-up rates for Ge-Si and Ge ad-dimers reach only about 30% of the value of the above hopping-down rate. In addition, the energy to remove the adatom from the step sink and move it to the lower terrace is rather smaller than 1 eV.[22] This indicates that the incorporation of Ge adatoms at the step sink is limited. In other words, the considerably high hopping rates, the much larger hopping-down rate than the hopping-up rate, and the limited incorporation rate of the ad-dimers at steps at 650° C will result in a fairly large flux of ad-dimers migrating downwards from the top terrace at the sidewall. Thus fewer Ge atoms will reside on the top terrace than corresponding to the nominally deposited value. Once the remaining Ge layer on the top terrace is below the critical thickness for island formation, no islands can be formed on the top terrace. On the other hand the aggregation of the Ge ad-dimers at the sidewalls facilitates island nucleation and formation there. This scenario qualitatively interprets the Ge island formation in samples X2, X7, X1 and X6, as shown in Fig. 1 and Figs. 3. (a)-(b).

This kinetic model can also explain the accumulation of GeSi at the trenches of the stripe pattern.[12] As a matter of fact, the thickness of the buffer layer, which is represented by the length of the white two-way arrows in Figs. 3 (a) and (b), on top of the terraces is smaller than that in the trenches. This thickness difference of the buffer layers is attributed to the migration of Si ad-dimers from the top terrace to the sidewall and from the sidewall to the bottom of the trenches. It confirms the above discussion.

**A. Effect of the growth temperature**

It is obvious that the position of Ge islands on the patterned stripes is related to the growth temperature since it significantly affects the migration of ad-dimers. This dependence of the position of Ge islands on the growth temperature is demonstrated by the AFM image of sample $X_T$, where 7 ML Ge are deposited at $600^o$ C after the Si buffer layer growth on the patterned stripes. As shown in Fig. 4, Ge islands are grown both on the top terrace and at the sidewalls. In addition, the Ge islands on the top terrace do not preferentially nucleate at the edge. Some small pyramid islands appear on the top terrace as well.

For the growth temperature of $600^o$ C and the mean top terrace width of 2220 Å, $t_{dmax}$ and $t_n$ are about 5.6$ms$ (for Ge-Si ad-dimer) and 8.1$ms$, as obtained by Eqs. (1) and (2), respectively. The difference of $t_n$ due to the differences of the top terrace widths for all these samples is rather small, however, $t_{dmax}$ at $600^o$ C is more than twice as long as at $650^o$ C. Consequently $t_{dmax}$ becomes comparable to $t_n$ which results in a large probability of the nucleation of triple-atom clusters, which will efficiently block the ad-dimers on the top terrace from reaching the edge at the beginning of each monolayer growth. Furthermore, the hop rate (downwards and upwards) of ad-dimers occurring at the step edge at $600^o$ C will be less than half of that at $650^o$ C. This leads to a smaller number of ad-dimers migrating from the edge of the top terrace to the sidewall. As a result, more ad-dimers migrating to the edge of the top terraces will reside at the edge or move back to the middle of the top terrace at $600^o$ C than at $650^o$ C, which will significantly increase the probability of the nucleation of a small static cluster at the edges or in the middle of the top terraces. Combining these two effects, most of the Ge atoms deposited on the top terrace can stay there to form islands. In addition, the nucleation of static triple-atom clusters on the top terrace takes place randomly, in the way of ad-dimers migrating to the edges. This may contribute to the irregular distribution of the final Ge islands on the top terrace, rather than to their preferential location along the edges, as shown in Fig. 4.

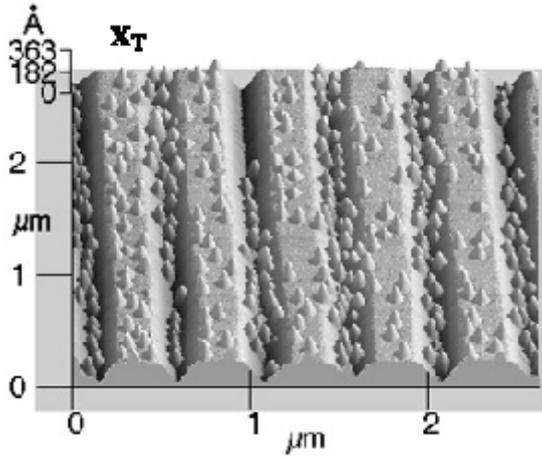

**Fig. 4** AFM image of sample $X_T$, where on stripe patterned Si substrates after a Si buffer layer 7 ML Ge are deposited at $600^o$ C.

Our observations on the preferred nucleation sites of islands are consistent with the enhancement of the mass transfer between the top terrace and the sidewalls at higher temperatures.[10] This influence of the growth temperature on ad-dimers (or adatoms) may contribute to the formation of Ge islands on the top terrace of the Si mesas at 570 $^o$C observed in Ref. 13. In addition, due to the absence of a Si buffer layer in the samples before the Ge deposition in Ref. 13, the roughness on the terraces[25] and of the step edges[26] will also contribute to the formation of Ge islands on the top and/or the edge of the top terrace by reducing the migration of ad-dimers.

**B. Effect of the step structure**

The position of Ge islands at the sidewalls of our samples, however, is random, as shown in Figs. 1 (a)-(c). The step structure at the sidewalls contributes to this randomness. It can be seen in the AFM micrographs that the sidewalls of the stripes generally form 'U'-like trenches. One cross-sectional height profile of the stripes in Fig. 1 (a) is shown in Fig. 5 (a). It is evident that the steepness of sidewall A (SWA) is larger than that of sidewall B (SWB), and an intersection is formed between SWA and SWB. The terrace width at the steeper SWA is narrower than that at SWB. Provided that step-step repulsion[19] is inversely proportional to the square of the terrace width at the sidewalls, the steps at SWA are not the favorable positions for Ge incorporation with respect to those at SWB. In addition, the narrower terrace width at SWA can contain fewer Ge ad-dimers. As a result, a large number of Ge ad-dimers will migrate to SWB from SWA. It can lead to a higher growth rate of SWB, especially in the region of SWB near the intersection where more ad-dimers from SWA will be incorporated in comparison to other regions of SWB. This preferential growth facilitates the island nucleation at those regions of SWB near intersections. The positions of these intersections at the sidewalls, which depend on the uniformity of the pattered stripes and the growth process, are random. Therefore, the Ge islands related to these intersections at the sidewalls are not laterally ordered, as shown in Figs. 1 (a)-(c).

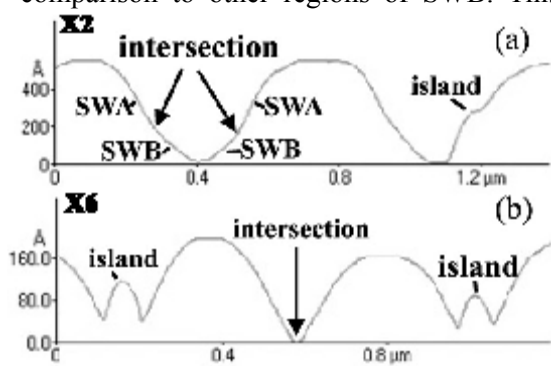 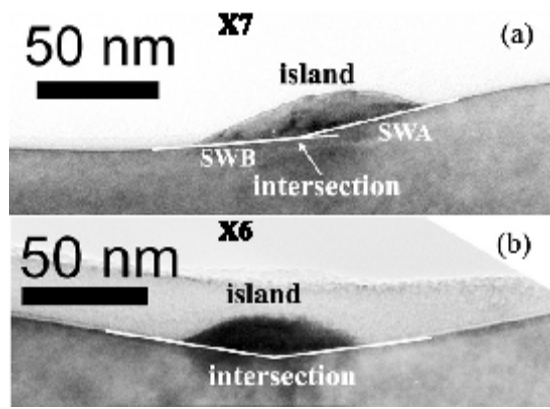

**Fig. 5** Cross-sectional height profile of the stripes in (a) sample X2, (b) sample X6. SWA and SWB in (a) represent two parts of the sidewall with different steepness, which results in a intersection, which promotes island nucleation. In (b) (note the different height scale) islands nucleate at the bottom of the V shaped trenches.

**Fig.6** Enlarged cross-sectional TEM images, (a) from sample X7, (b) from sample X6. The white lines denote the sidewalls in the vicinity of the islands, which result in intersections under the islands.

On the other hand, the cross-sectional height profile of the stripes in Fig 1 (d) demonstrates that the sidewalls of the stripes in sample X6 form shallow 'V'-like trenches, as shown in Fig. 5 (b). There is only one intersection at the bottom of the trenches. Due to the small number of the steps at the sidewalls, a large number of Ge ad-dimers can migrate to this intersection site before they are incorporated at the sidewalls. This aggregation of the Ge ad-dimers at the intersection promotes the formation of the Ge islands. In addition, because the bottom of the shallow trench is always linear, *one-dimensionally ordered islands* can be readily achieved, as shown in Fig. 1(d). The enlarged cross-sectional TEM images of samples X7 and X6 in Fig. 6 (a) and (b) also demonstrates the relation between the positions of the islands and the intersections at the sidewalls or the bottom of the trenches, respectively.

In the above discussion, the ad-dimer migration parallel to the stripe, the meandering or kinks of the steps and the mixture of single-layer step was not considered. Although this local step structure essentially has no effect on the nucleation of islands on the top terrace, it will affect the nucleation of islands at sidewalls.[8]

### C. Effect of external strain fields

The external strain field produced by the SiGe buffer layer in samples X3 and X4 influences the migration of the ad-dimers and the local critical thickness for the 2D-3D transition, which leads to the preferential formation of Ge islands on the top terrace of the stripes. The upper curve in Fig. 7 shows the distributions of lateral strain for sample X3 (one period of the stripe structure), where $\varepsilon_\parallel = (a_\parallel - a_{Si})/a_{Si}$ ($a_\parallel$ is the lateral lattice constant of epilayer, $a_{Si}$ is the lattice constant of the Si bulk). The inhomogeneous strain distribution was calculated by the finite element method (FEM) using the Patran program package. The schematic cross-sectional image of one stripe before Ge deposition, derived from the AFM image and the TEM image, is also shown in the lower part of Fig. 7. Apparently, the misfit strain of the subsequently grown Ge layer at the corner between the top terrace and the sidewall can be partially relaxed. Moreover, a strain gradient is introduced there, which is known to change the migration of the adatoms.[27] Qualitatively, the strain gradient around the top corner results in an increase of the activation barrier for Ge-Si (or Ge) ad-dimers to hop down, and even results in a higher activation barrier to hop down than that to hop up at the steps of the sidewall above point A in Fig. 7. This means that less ad-dimers on the top terrace can migrate down to the sidewall, and even a net flux of ad-dimers from the sidewall above the point A to the top terrace can take place in contrast to the case where no SiGe buffer layer is grown. As a result, enough Ge atoms for island formation can reside on the top terrace. In addition, the strain status at the center of the top terrace is compressive and at the edge it is tensile, as shown in Fig. 7. This means that the critical thickness for the 2D-3D transition at the center will be smaller than that at the edges. Mui et al.[7] have demonstrated that the formation of islands sensitively depends on the thickness of the deposited strained layer with respect to the critical thickness. Under the assumption of the SK growth mode, the difference of the critical thickness at the center and at the edge might be large enough to lead to an island nucleation first at the center, rather than at the edges of the stripes in sample X3, as shown in Fig. 2 (b). This difference of the local critical thicknesses due to the strained layer may also explain the previously reported formation of InAs islands at the center of GaAs mesas.[28] For sample X4, there is a similar strain distribution at the surface of the stripes. This external strain field, together with the limited top terrace width, leads to a linear row of Ge islands on the top terrace, as shown in Fig. 2 (a).

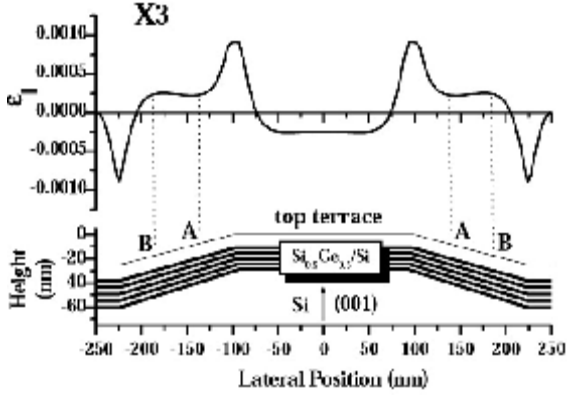

**Fig. 7** Lateral strain distribution ( in plane strain tensor component) at the surface of the stripes (one period) induced by the buried Si/SiGe superlattice buffer layer. The schematic cross-sectional image of one stripe before the Ge deposition is displayed at the bottom. The heights and widths are derived from the AFM and the TEM images of sample X3.

However, the effect of the external strain field to the migration of Ge (or Ge-Si) ad-dimers at the sidewalls between the point A and point B, as shown in Fig. 7, can be neglected because the strain gradient is nearly zero in this range. Therefore, Ge-Si (or Ge) ad-dimers in this range can still migrate downward at the sidewall as in the absence of an external strain field. When the thickness of the Ge layer at the bottom part of the sidewall is beyond the critical value, Ge islands are formed there as well, as denoted by the large black arrow in Fig. 2 and Fig. 3 (c). The larger number of Ge islands near the bottom corner in the Fig. 2 (a) than in the Fig. 2 (b) is mainly attributed to the larger size of the sidewall, which can provide more Ge atoms for the formation of islands.

Ge islands have also been found to grow on the top terrace of Si mesas or stripes[9,10] selectively grown in windows within a $SiO_2$ layer. In these cases, two facts should be taken into account. The first one is the existence of the $SiO_2$ layer around the mesas or stripes during the growth. When the $SiO_2$ layer is on the top terrace, the adatoms will migrate down at the sidewalls to the corners between the sidewalls and the bottom terraces to form quantum wires or islands.[29] However, the adatoms migrate up at the sidewalls to the top terrace when the $SiO_2$ layer is around the bottom of the growing stripes or mesas.[30] This influence is attributed to the external strain field induced by the $SiO_2$ layer, which alternates the migration of adatoms similar to the strain field induced by a strained SiGe buffer layer, as discussed above. The second important fact is the existence of hydrogen at the growing surface in gas source MBE in Refs. 9 and 10 , which also affects the migration of the adatoms.[31]

## IV. SUMMARY

In summary, we analyzed the preferential positioning of islands grown at 650°C on patterned Si stripes with and without a strained SiGe buffer layer. The surface morphology was investigated by AFM, and in addition cross-sectional TEM images were made. The Ge islands do not preferentially grow at the edges of top terraces, where it would be energetically favourable. We propose that the growth kinetics primarily affect the nucleation and the formation of the Ge islands on patterned Si substrates. The preferential positioning of the islands is also related to the step structure and to external strain fields.

By decreasing the period of the patterned Si stripes and optimising the buffer layer growth, shallow 'V'-like trenches can result which will lead to one-dimensional ordered islands with a high density, along these trenches for a growth temperature of 650°C. On the other hand ordered arrays of islands on the top terraces are found, if strained SiGe buffer layers are deposited prior to the Ge island deposition. Lower growth temperatures can result in Ge islands nucleating both on the top terraces and the sidewalls of the patterned stripes.

**Acknowledgements**
We thank P. Mayer and A. Hesse for their help with FEM calculations. This work was supported by the FWF, Vienna (project No 14684), the EC project ECOPRO.